\documentclass{article}

\usepackage{PRIMEarxiv}

\usepackage[utf8]{inputenc} 
\usepackage[T1]{fontenc}    
\usepackage{hyperref}       
\usepackage{url}            
\usepackage{booktabs}       
\usepackage{amsfonts}       
\usepackage{nicefrac}       
\usepackage{microtype}      
\usepackage{lipsum}
\usepackage{fancyhdr}       
\usepackage{graphicx}       
\graphicspath{{media/}}     
\usepackage{authblk}

\usepackage{bbm, dsfont}
\title{Anatomy of Perturbed Traffic Networks during Urban Flooding
}

\author[1]{Akhil Anil Rajput \thanks{akhil.rajput@tamu.edu}}
\author[2]{Sanjay Nayak \thanks{sanjaynayak@tamu.edu}}
\author[3]{Shangjia Dong \thanks{sjdong@udel.edu}}
\author[1*]{Ali Mostafavi\thanks{amostafavi@civil.tamu.edu}}
\affil[1]{Zachry Department of Civil and Environmental Engineering, Texas A\&M University, College Station, TX 77840, USA}
\affil[2]{Department of Computer Science and Engineering, Texas A\&M University, College Station, TX 77843, USA}
\affil[3]{Department of Civil and Environmental Engineering, University of Delaware, Newark, DE 19716, USA}
\affil[*]{Corresponding author}

\begin{document}
\maketitle




\section{Abstract}

Urban flooding disrupts traffic networks, affecting the mobility and disrupting access of residents. Since flooding events are predicted to increase due to climate change, and given the criticality of traffic networks, understanding the flood-caused disruption of traffic networks is critical to improve emergency planning and city resilience. Leveraging high-resolution traffic network data from a major flood event and advanced high-order network analysis, this study reveals characteristics of the anatomy of perturbed traffic networks. First, the findings show network-wide persistent increased travel times could last for weeks after flood water has receded, even after modest flood failure. During disaster event period, modest flooding of 1.3\% road segments caused 8\% temporal expansion of the entire traffic network. The results also show distant trips would experience a greater percentage increase in travel time. Also, the extent of increase in travel time does not decay with distance from inundated areas, suggesting that the spatial reach of flood impacts extends beyond flooded areas. Departing from the existing literature, which primarily informs about physical vulnerability in road networks to floods, the findings of this study provide important novel understanding of floods impacts on the functioning of traffic networks in terms of travel time and traffic network geometry. The persistent travel time increase in the entire network can translate to significant social and economic impacts in terms of user costs, additional CO$_2$ emissions, and lost productivity. Due to the importance of traffic networks functioning for the operation of cities, the results have significance implications for city managers, transportation planners, and emergency managers to deal with the impacts of urban flooding.

\section{Introduction}

Transportation networks connect populations and services ~\cite{fema_2020}. The stability of a transportation network is challenged by flood hazards ~\cite{PREGNOLATO201767}  which can trigger compound physical and functional failure that result in network connectivity loss \cite{dong2022modest}. Community recovery is further impacted when the access to critical facilities such as fire stations, shelters and hospitals are disrupted ~\cite{fan_jiang_lee_mostafavi_2022, YUAN2022101870}. The extent of impact is expected to increase due to climate change ~\cite{WASKO2021126994, Ghanbari_climate}. Researchers have sought to understand how floods disrupt transportation networks \cite{dong2022modest, wang2019local} to improve infrastructure resilience planning \cite{esmalian2022operationalizing}. Existing studies, however, focus mainly on either physical road network topology during disruptions \cite{wang2019local, BAGLOEE201760, MATTSSON201516} or on transportation functionality in normal conditions without disruption \cite{hamedmoghadam2021percolation, li2015percolation}. Little attention is devoted to the time-varying link functionality in transportation networks. The flow of traffic through the network, as well as network connectivity, is essential to functioning of a community. But the flood impact on traffic networks is not yet fully understood.  

The use of percolation methods ~\cite{stauffer_aharony_2018} to analyze physical road networks provides limited insights regarding floods' impacts on transportation systems. Although such measures adequately quantify the extent of the impact on road networks, they give little to no insights into how travel is impacted in the city. Percolation-based analysis informs about the physical vulnerability of networks but does not inform about impacts on transportation system functioning. One key indicator of the functioning of traffic networks is travel time. Some studies have tried to address this using the percolation approach ~\cite{SOHOUENOU2021102672, doi:10.1126/sciadv.1701079} but there is limited research on the understanding of traffic networks under natural disasters such as flooding. However, little is known about the extent to which floods perturb travel time in traffic networks and whether the impacts on traffic networks would be local to flooded areas or affect the more significant part of the network. Or how long would the travel time impacts persist in the network after the flood recedes? Therefore, percolation analysis does not fully capture real-world networks' temporal dynamics and spatiality. Recent studies have shown the significance of understanding the geometric properties of spatial infrastructure networks such as road networks ~\cite{https://doi.org/10.1002/geo2.95, ijgi8070304, urbansci6010011}. However, when the flow dynamics on the network are involved, we need to derive new metrics to understand the network resilience properties. Traffic networks are spatially embedded in cities and communities, and their link dynamic varies temporally ~\cite{batty_axhausen_giannotti_pozdnoukhov_bazzani_wachowicz_ouzounis_portugali_2012, doi:10.1177/2399808319837982}. We have learned the impact of floods on the spatial geometry of the physical road network, but how the geometry of the traffic network changes in the time domain has yet to be fully understood. For example, when road inundations and heavy congestion increase the travel time between two spatial nodes (i.e., road junction), this is equivalent to the two spatial nodes becoming more distant from each other. Hence, the temporal geometry of spatially-embedded traffic networks would change.

To this end, the goals of this research are to assess (1) the extent to which floods perturb travel time in traffic networks, (2) whether the impacts on traffic networks would be isolated to flooded areas or would affect a larger part of the network, and (3) length of time that the impacts on travel time persist in the network after the flood recedes.Traffic networks are defined as representations of a network of roads with time-varying functionality. To characterize the anatomy of perturbed traffic networks during floods, we adopted two novel geometric properties of the dynamic traffic network (Figure 1): (1) network expansion and (2) simplicial complex change. Network expansion refers to the extent to which travel time between the node pairs (road junction pairs) in the networks increases due to perturbations. In flooding, road inundations and congestion would increase travel time between node pairs and hence, cause a virtual expansion in traffic network topology. Simplicial complexes represent the topological geometry of networks ~\cite{Torres_2020}. They captures higher-order topological changes in traffic networks during flooding. Hence, the examination of changes in the higher-order traffic networks with time-varying link functionality can provide a better understanding of the perturbed traffic networks during floods. Both network expansion and simplicial complex change simultaneously capture the effects of road inundations and congestion caused by flooding, providing a more complete understanding and quantification of flood impact on traffic networks. 

Using high-resolution empirical traffic data from Harris County, Texas, collected during Hurricane Harvey (2017), We first examined the average shortest travel time between node pairs (road intersections) during normal status and during flood-disrupted states to quantify the extent to which flooding expands travel time between node pairs, and to infer the virtual expansion of the traffic network. Second, we examined the Betti number at different filtration levels in traffic networks, as fluctuations in the Betti number reflect the traffic network simplicial complex change. Fluctuations in the Betti number expose the characteristics of higher-order network changes and reveal the extent of changes in traffic network topological features when flooding causes direct (road inundation) and indirect (congestion) perturbations. Using this travel time-based characterization of the traffic network, the findings of this study move us closer to a more complete understanding of the impacts of flooding on transportation systems and the functioning of cities.  

\begin{figure*}[!ht]
\centering
\includegraphics[width=0.95\textwidth]{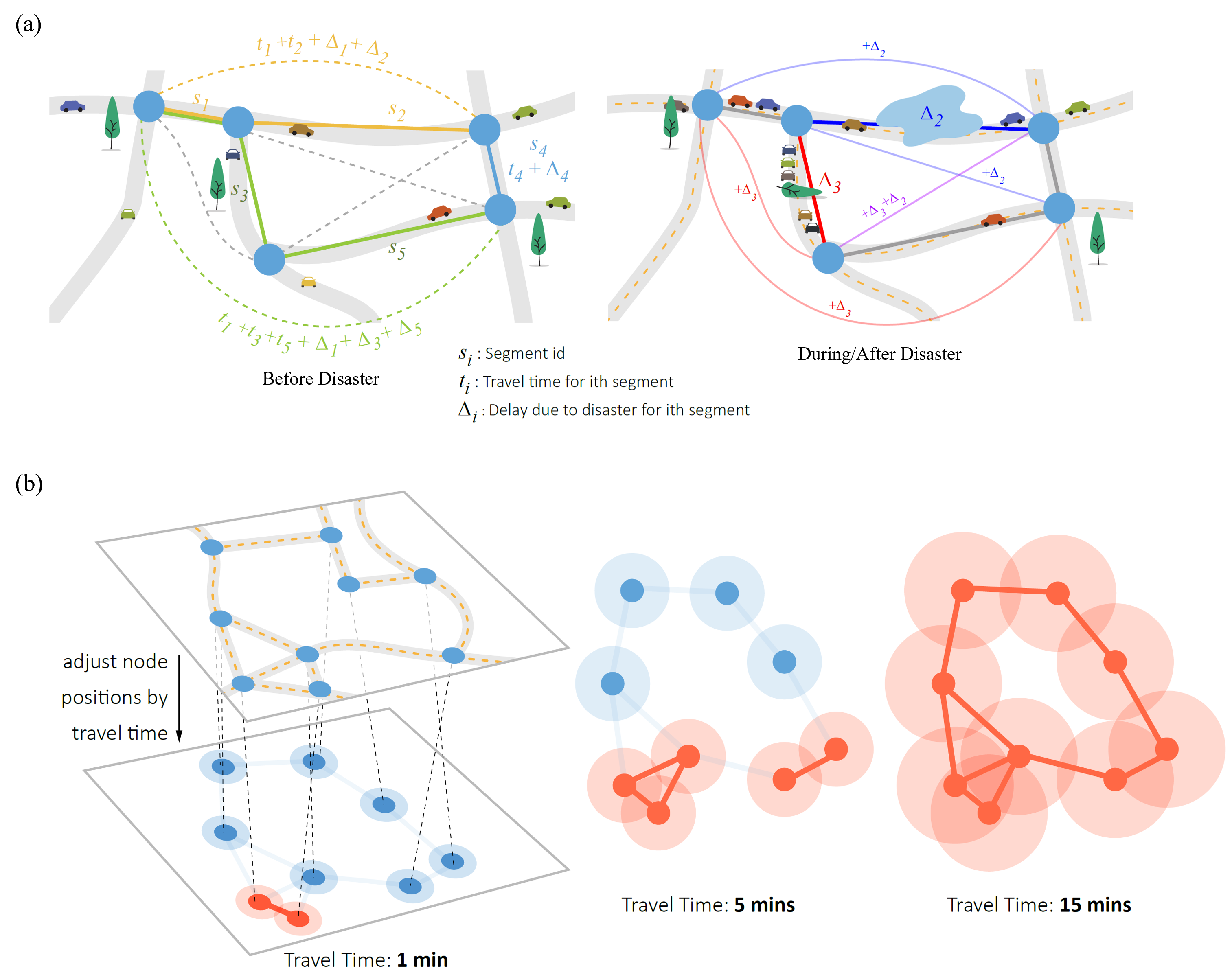}
\caption{Conceptual illustration of the analysis performed. (a) Illustration of pair wise travel time and changes in temporal links due to perturbations. (b) Connected component framework at different Filtration levels. Each filtration level corresponds to a travel time, within which the nodes (road junctions) in the network are connected, although they may not have a direct link connectivity. Metric for the number of connected components at each filtration level represents the most basic higher order network analysis metric.}
\label{fig:multilayer}
\end{figure*}

\section{Results}

\subsection{Persistent travel time increase and temporal expansion in the entire traffic network}

\begin{figure*}[!ht]
\centering
\includegraphics[width=0.75\textwidth]{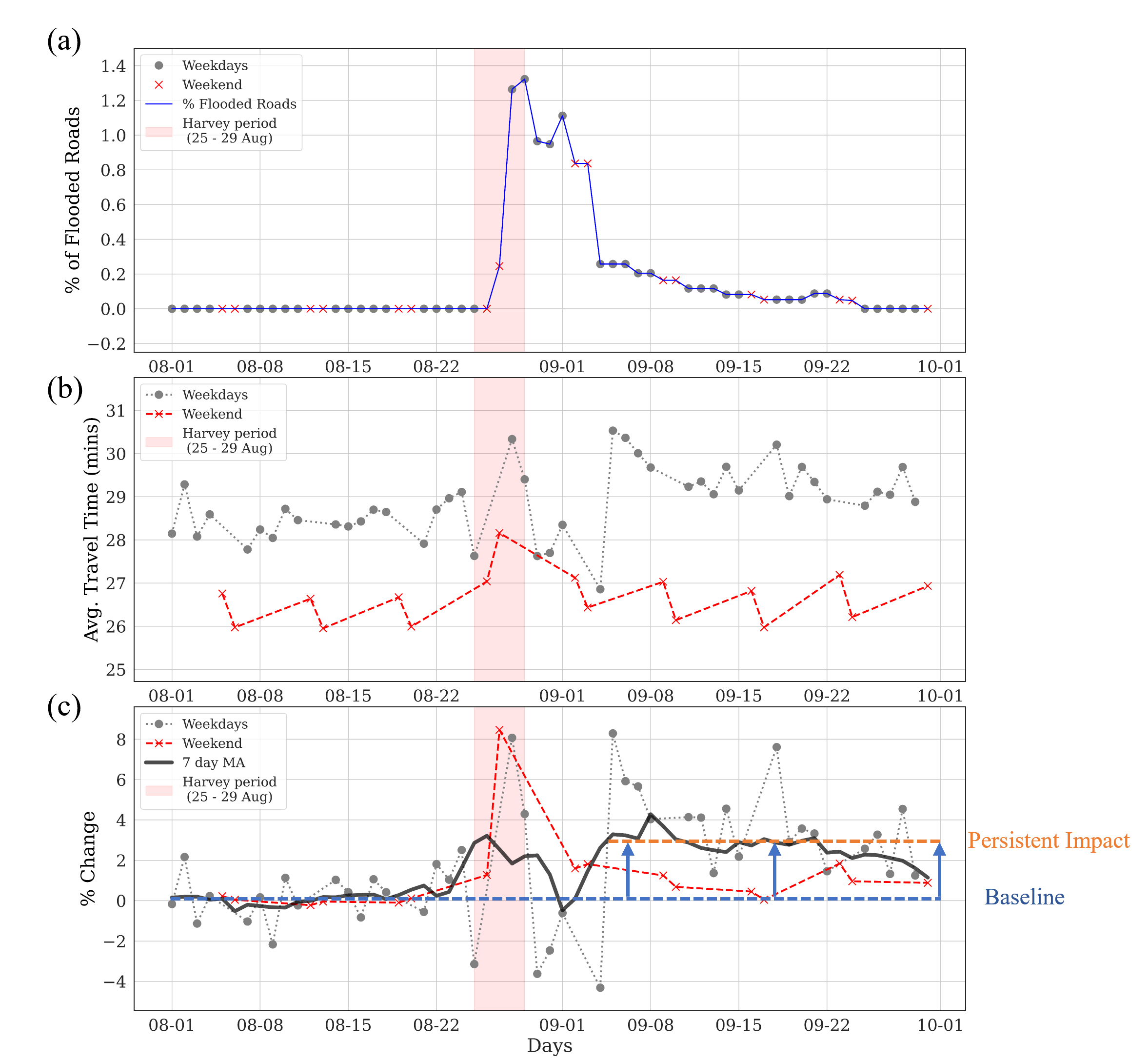}
\caption{Percentage change in travel time between every pair of road junction in Harris county during, before and after Harvey. (a) Percentage of flooded edges as a function of time. After Harvey leaves Houston, further flooding happens due to reservoir water release leading to addition road closures immediately after Harvey period. (b) Average daily travel time between pairwise junctions in minutes. Weekdays and weekends show distinct trip patterns due to difference in lifestyle and lack of work-related trips on weekends. Work-related trips increase average traffic and leads to higher travel times compared to weekends. (c) Change in the pairwise travel time compared to the baseline values. The same day of the week is compared across weeks to account for differences in movements for different days of the week. For a modest 1.3\% of flooded roads, 8\% increase in travel time is observed for the entire network on August 27 and 28. After roads recover from inundations, with only 0.25\% roads remain perturbed, sustained impact is observed for travel times. The entire network experiences an average of 3\% increase of travel time among node pairs even a month after Harvey landfall.}
\label{fig:avg_change}
\end{figure*}

Hurricane Harvey made landfall in Harris County on August 25, 2017, and significantly disturbed the traffic network. In the average network-wide pairwise travel time at different phases of the Harvey (Fig \ref{fig:avg_change}), we see that the travel time first decreased at the time of Harvey landfall (August 25) due to residents sheltering in place, thus reducing travel demand, in anticipation of Harvey's landing. Travel time then increases (August 28) due to the increase in flooding-induced road closures. The travel time drops again between August 28 through 30, as flooding is receding but residents still have not started traveling. By September (Fig \ref{fig:avg_change} a), the number of flooded/perturbed road segments decreased from 1.3\% (August 28 and 29) to 0.25\%. The travel time slowly recovers as flooded roads become available and debris is removed. By September5, road conditions are largely improved and travel demand is headed towards normalcy. Longer travel times due to congestion have persistent impact on the travel time for the entire network for perturbed road segments that account for less than 0.25\% of road segments. 

A small fraction of road closures impacts on the entire traffic network. According to Fig ~\ref{fig:avg_change} (a), at peak inundation, 1.3\% of perturbed roads contributes to an average of 8\% of increases in travel time in the entire network, which is equivalent to the network being expanded by 8\% (every junction pair gets more distant from each other by 8\% travel time). However, it is interesting to note that both 1.3\% (August 28 and 29) and 0.25\% (Sep 4) of flooding-induced road closure can result in an 8\% of increase in travel time. This affirms two findings: (1) the location of flooding is important. When a small number of critical roads are perturbed, the impact is as extensive as the disruption of multiple ordinary roads; (2) accounting for disturbed travel demand due to flooding is a factor in assessing the impacts on the traffic network in terms of travel time.During Hurricane Harvey, many roads were closed, thus increasing overall travel time. In the post-Harvey period, travel demand picked up, and thus more congestion on the road (while fewer road segments were perturbed). With a small proportion of road closure remaining, the compound flooding and congestion impacts led to an increase in the travel time of 8\% on September 4. \cite{dong2022modest}. 

Flooding affects traffic networks differently during weekdays and weekends. As shown in Fig ~\ref{fig:avg_change} (b and c), Travel times during weekdays and weekends were both disturbed by the flooding; however, weekend travel time quickly recovered to the pre-Harvey level, while weekday travel sustained the average 3\% of travel time increase even one month after the Harvey. This persistent travel time increase during weekdays can be attributed to the weekday commute demand change in the aftermath of flooding. Weekend travel needs and schedules tend to be flexible, thus travel time returned to normal during weekends more quickly. The 3\% persistent travel time increase in the entire network during weekdays can translate to significant social and economic impacts in terms of user costs, additional $ CO_{2} $ emissions, and lost productivity.   

\subsection{Flooding disproportionately prolongs long-duration travel}

\begin{figure*}[!ht]
\centering
\includegraphics[width=1\textwidth]{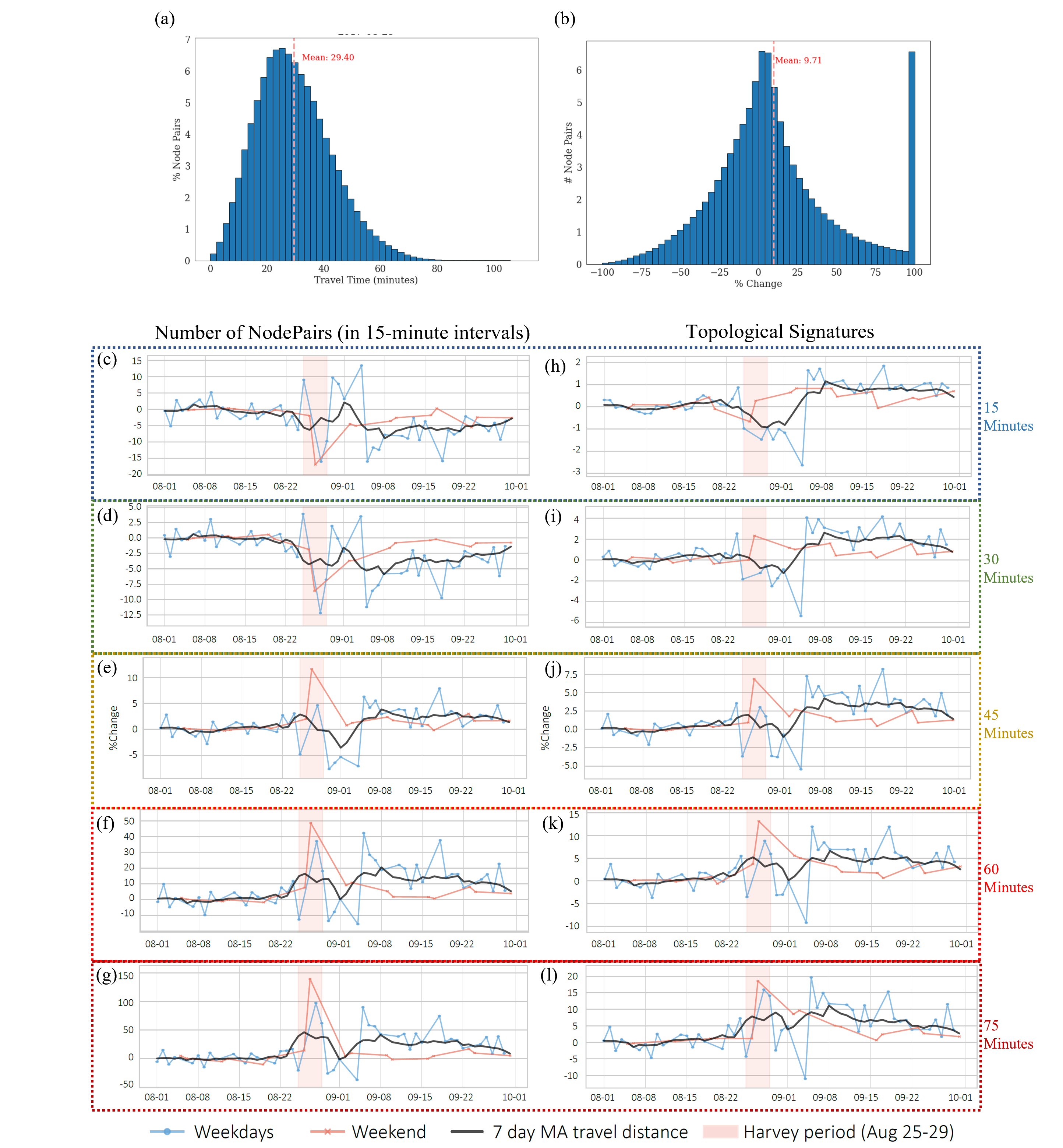}
\caption{Perturbation characteristics of long and short trips in traffic network. (a) Distribution of the number of node pairs (junction-to junction-travel pairs) versus travel time for August 29. Distribution for all days follows a bell curve for and has a mean value of around 30 minutes. (b) Proportion of the number of node pairs corresponding to travel time percent change due to Harvey on August 29. Change of more than 100\% was considered as 100\% for better visual clarity. (c-g) shows change in the trips for intervals of 0–15, 15–30, 30–45, 45–60, 60–75 minutes intervals respectively. (h-l) shows change in the number of connected components at thresholds of 15, 30, 45, 60 and 75 minutes, respectively. Longer trips show higher impact due to traffic disruptions as impacts compound for longer commutes. Travel time and topology-based impact assessment show different characteristics of disruption and recovery for different time intervals. Post-disaster sustained impact is seen in every time range in both assessments.}
\label{fig:travel_betty_change}
\end{figure*}

For the shortest path between pair-wise road junctions (also called node pairs), travel time is computed by summing the time contribution of perturbations or disturbance of each road segment. This method has a compounding effect on the overall time for movement from one location to another. We evaluated the impact of flooding on trip ranges by segregating trips into  15-minute intervals: trips of less than 15 minutes, between 15 up to 30 minutes, 30 up to 45 minutes and so on, the last interval being 60 to 75 minutes. We then count the number of junction pairs that fall within each travel time range and compare them with baseline. This parameter provides insights about the disproportionate impact on junctions within different temporal proximity. This interval classification accounts for more than 99.9\% of the trips on non-impact days; therefore, represents the entire traffic network reliably. The distribution of travel time versus fraction of node pairs (Fig ~\ref{fig:travel_betty_change} (a)) suggests that, during peak inundation, travel time follows long-tailed distribution. Changes in travel time of pairwise junctions on August 29 (Fig. ~\ref{fig:travel_betty_change} b) suggests that while peak inundation induces an overall increase in travel time, due to disconnection of some junctions from main network due to closure related to inundation and road damage. Nevertheless, a large proportion of road segments show a decrease in travel time.

The results shown in Fig ~\ref{fig:travel_betty_change} (c-g) provide insights on the extent of change and impact for travel time ranges. On average, trips with a mean travel time of less than 30 minutes show a decrease of about 5 to 10\%. This is because trips of shorter travel time, thus fewer road segments, are less subject to  compounding effects. On the other hand, the extent of increase in trips of greater than 60 minutes is about 50\% on average and increases to about 140\% during peak impact day. On average, the extent of impact increases as travel times increase between road junction. Thus, the impact on travel times due to urban flooding is directly proportional to the distance between the places. 

We also see a sharp decline in shorter travel times immediately after Harvey, as there is less traffic on road. As evacuated residents return  and city recovers to normalcy, travel times increase exponentially, reaching the same level as that during Hurricane Harvey. This is due to the fact that some road segments are still littered with debris or closures, but still must to cater to  high demand. It is worth noting that the average levels of change for all time intervals reach almost the same level as that observed during Harvey. This indicates that although the actual landfall lasted only for a couple of days its impact was observed at virtually the same intensity on an average until the end of September. 

To examine the impact of flooding on higher-order network measures, we use a topology based measure, Betti-0, that computes the number of connected components in a network at different travel time thresholds. In the context of a traffic network, Betti number at a threshold of 15 minutes ($\epsilon_1 = 15$) would look at the number of connected network components when road junctions within a 15-minute proximity are merged and considered as one component. At the initial thresholds of travel time, there are multiple pockets of such connected components, since not all junctions are reachable by one another given the threshold. Hence, a number of clusters get formed.
These clusters represent places of closest proximity in terms of travel time. For simplicity we focus only on five thresholds, $\epsilon_i, i = 15, 30, 45, 60, 75$. The results of the percentage change in Betti numbers at different days for these five filtration values are shown in Fig. \ref{fig:travel_betty_change} (h-l) . 

The results indicate that changes in the topological features in the network follow distinct patterns. For features within 15- and 30-minute thresholds, we first see a decrease in the number of such connected components, then an increase, followed by slight decrease before reaching at equilibrium. For other filtration levels, we see an increasing trend till the second day of Harvey (August 26)  then a decreasing trend till immediately after Harvey (September 1). There is then an increase in the connected components at the respective travel time thresholds that stagnates at an higher or similar level, as observed during Harvey. The percentage change in the number of connected components within different time intervals is less than the changes observed in the variation in number of junctions connected at different thresholds (Fig. \ref{fig:travel_betty_change} h-l). This result indicates that higher filtration levels (associated with connectivity of junctions with longer travel times) show more sensitivity to changes in the network due to flooding. This result confirms the earlier results regarding the greater sensitivity of longer travels to flooding impacts.

The reduced number of the connected components during Harvey for shorter trips is a result of decreased travel time due to less traffic, making a greater number of nodes reachable within a time threshold as compared to pre-Harvey conditions. In contrast, there was an increase in the number of connected components for longer trips during Harvey, implying a lower reachability given the same time window, as if flooding caused an invisible temporal expansion of the entire road network of the city. This temporal expansion of the traffic network influences the higher-order structures in the network and makes more junctions reachable for shorter trips in terms of travel time and less junctions for longer duration trips. The differences in travel time change for various filtration levels reveals that floods affect travel durations disproportionately, putting longer-distance travels in jeopardy. Coupled with critical service needs and accessibility, such impact disparity can further exacerbate the community vulnerability \cite{dong2020integrated}.

\subsection{The extent of travel time change does not decay with distance from inundated areas}

\begin{figure*}[!ht]
\centering
\includegraphics[width=1\textwidth]{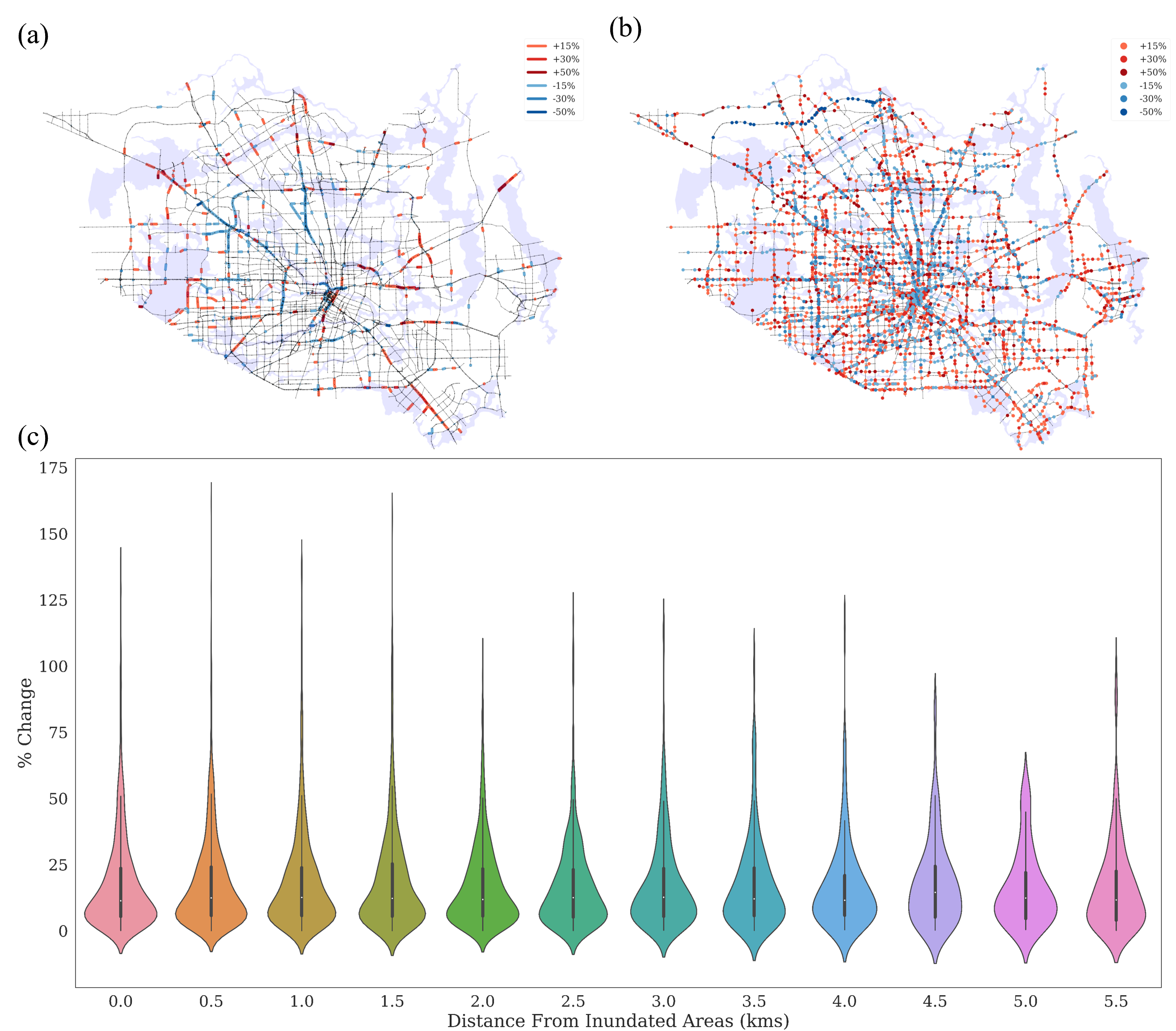}
\caption{Spatial impact of floods on traffic network on August 29. (a) Percent change in travel time for each road segment along with flooded regions for that day. (b) Average impact in terms of percent change in travel time at each node (road junction) when trips are considered to every other node. (c) Changes in travel time as a function of distance from flooded region. Change for regions within inundated region is shown by Violin plot at 0 distance; plot at 5.5 km distance presents the aggregated changes for all road junctions more than 5 km away, which accounts for less than 1\% of the road junctions. There is no decay in change of travel time with respect to distance to inundated areas. Areas far from the flooded regions also show same extent of travel time change on an average.}
\label{fig:Spatial_Change}
\end{figure*}

We evaluated the spatial patterns of travel time changes with respect to proximity to inundated areas. We spatially visualized the junctions that show an overall magnitude of change of more than 15\%. Here we assess the impact by aggregating the travel times from one junction to every other junction and calculate the average change in travel time at a junction. We compared this result with road segments having an average travel time change of more than 15\% to evaluate if they exhibit spatial colocation. Fig. \ref{fig:Spatial_Change} (a) and (b) show spatial occurrence of the specified junctions and road segment, respectively, on peak flooding day, August 29. The effect of perturbation in the filtered road segments can be be seen in Fig. \ref{fig:spatial_change_disaster} (c) which corresponds to August 27, two days after landfall of Harvey in Harris County; Fig. \ref{fig:spatial_change_disaster} (d) shows the road segments that experienced an increase in travel times due to flooding for the same day. Although the road segments on the major highways show increased travel times, the effect can be seen over the entire network. Most of the junctions show an increase in travel time of more than 15\%, with some showing more than 50\% increase. Similar insights can be obtained by comparing the results for August 28 and 29.  

When Harvey dissipated in the Houston area August 30 and 31, the majority of junctions experienced a reduction in travel time, consistent with the results obtained from comparing average travel times in the overall network. Although some regions, such as southwest Harris County, retained road segments with increased travel times, the effect cannot be seen locally or throughout entire network. But a week later, the increase of travel time resolved in Southwest area, despite roads unaffected by flooding showing increased travel time. This result provides  evidence that, although flood-related impacts on road network is local, the spatial reach of flooding on the overall travel time and connectivity is extended beyond inundated areas. This spatial reach does not decay with distance from inundated areas.  

Further investigation of the absence of spatial decay evaluated the change in travel time with distance from flooding (Fig. \ref{fig:Spatial_Change} c). The median change in the travel time at every junction, considering travel to every other junction, shows a similar pattern irrespective of the distance from flooded region. Additionally, junctions in flooded areas have the same change as those outside inundated areas, demonstrating that flooding affects the entire traffic network irrespective of direct proximity to flooded regions. We do not, therefore, observe any decay in flooding impact on travel time with distance from flooded regions. While Li et al., 2022  ~\cite{doi:10.1073/pnas.2203042119} show that mobility exhibits spatio-temporal decay from crisis locations when observed at county, state, and country resolution, our analysis at a much finer resolution does not indicate the presence of spatial decay in the impacts flood on traffic networks.

\section{Discussion}

This study examines the virtual expansion of traffic networks during flooding by considering flood impact on travel time. The results reveal three novel properties of perturbed traffic networks caused by urban flooding: (1) persistent entire network travel time increase, (2) long-tail effects on long-travel distance travels, and (3) absence of spatial decay in travel time changes with distance from inundated areas. Specifically, the results show that 1.3\% of flooded roads during Hurricane Harvey in 2017 were responsible for an 8\% increase in overall travel time throughout the network. The impacts of flooding on traffic networks persists for several weeks after inundation receded. Furthermore, such impact on travel time is not homogeneous but affects longer trips (i.e., 45–60 minutes) more strongly than shorter ones (i.e., less than 15 minutes). Such a heterogeneous impact of flooding on travel times is a  factor to be considered in disaster traffic management to maintain a community's access to critical services. Investigation of high-dimensional features using the Betti number reveals that flooding imposes an impact on traffic congestion post-disaster, which can be as high as that observed during peak inundations. Although flood disruption on road segments is localized, the generated impact is diffused throughout the network, suggesting that the impact on the travel time in a city is invariant of the location of disruption. Moreover, the impact is sustained even one month after flooding and causes a 3\% expansion of traffic network for a fraction of unrestored road segments. 

This study has multiple contributions: first, the findings reveal the impact of floods on travel times in urban traffic networks. Prior studies  focused  on vulnerability of physical roads; our understanding of the perturbed functioning of traffic networks during floods was limited. Given the importance of traffic network function in terms of travel time, the findings of this study can inform city managers, transportation planners, and emergency responders about the persistent and entire network impacts of local floods which are expected to grow with climate change impacts. The persistent travel time increase in the entire network can translate to significant social and economic impacts in terms of user costs, additional $ CO_2 $ emissions, and lost productivity. Second, this study employed topological network measures and higher-order network analysis to capture both temporal dynamics and spatiality of traffic networks. The prior studies on urban networks were primarily based on percolation analysis and were not able to capture temporal dynamics of links functionality as well as the spatiality of real world networks. The approach used in this study can be employed for assessing the resilience of other spatially embedded and temporally dynamic networks, such as power grid networks. Third, unlike the majority of studies which use location-based human mobility data for analyzing origin-destination trip fluctuations in floods and other crises, this study dissected fine-resolution link-level travel time data to analyze the perturbed dynamics of traffic networks. The fluctuations in human mobility do not fully capture the functionality of traffic networks in terms of travel time (the primary function of transportation networks). The number of trips might return to normal, but the travel time between junctions may stay elevated for a longer duration. Hence, the novel insights obtained from this study move us closer to a better understanding of the impacts of floods on urban traffic networks.

Urban flooding is a threat to large metropolitan cities, and the frequency of floods is expected to increase with climate change. The revealed persistent and network-wide impact of floods and their heterogeneous impacts on trips of varying lengths provide  evidence for planners and emergency officials to effectively manage the city traffic during urban floods to ensure proper functioning of cities. The pairwise junction travel assessment method and higher order analysis employed in this study capture both temporal dynamics and spatiality of traffic networks. The findings of this study would be generalizable to other cities, and flood events since traffic and mobility networks show similar characteristics in cities thought the world ~\cite{noulas_scellato_lambiotte_pontil_mascolo_2012, chan_donner_lämmer_2011}. Therefore, they are likely to exhibit similar patterns of disruptions and recovery during disasters. Moreover, This method can be transferred to other spatially embedded and dynamic temporal networks and disaster scenarios, such as the power grid during storm events. Application of these methods on traffic networks showed that on average, localized impact has same effect on travel times away from disrupted regions as those within and in the nearest proximity to disruptions locations. This study complements existing location-based human mobility studies by dissecting fine-resolution link-level travel time data to analyze the anatomy of flood-perturbed traffic networks. We reveal that although the total number of trips might return to normal after flooding, the travel time between junctions can persist for a longer duration. 

\section{Data and Methods}

\begin{figure*}[!ht]
\centering
\includegraphics[width=1\textwidth]{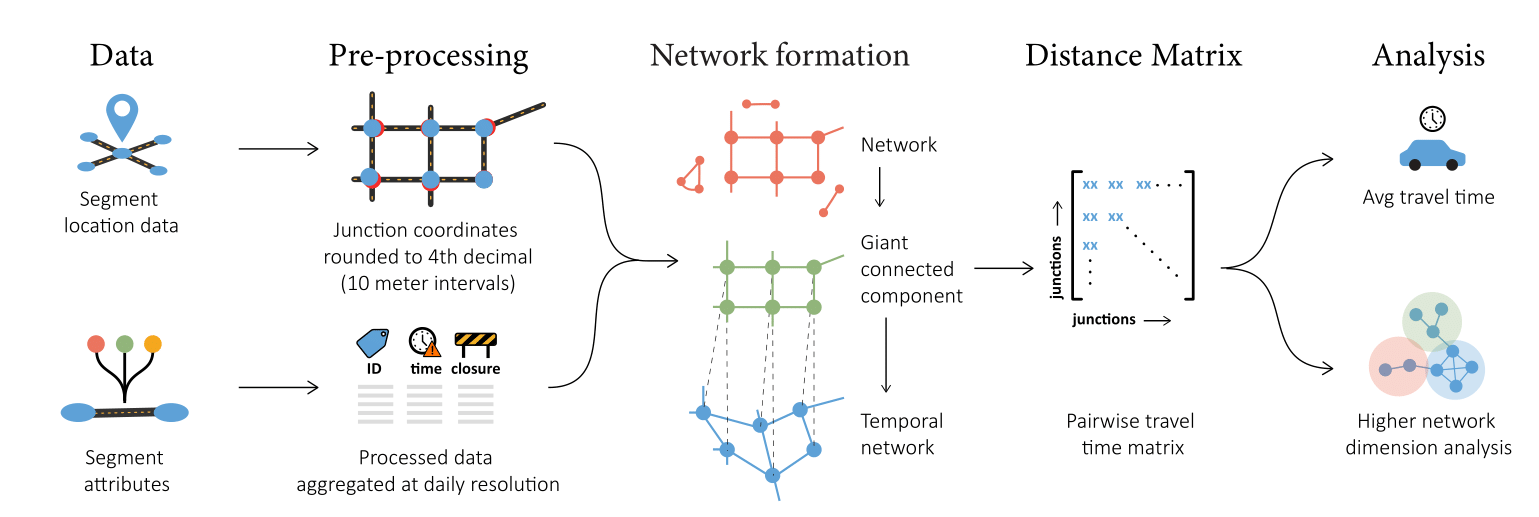}
\caption{Framework used in this study to evaluate the dynamics of change in  spatio-temporal traffic network.}
\label{fig:methods}
\end{figure*}

\begin{figure*}[!ht]
\centering
\includegraphics[width=0.8\textwidth]{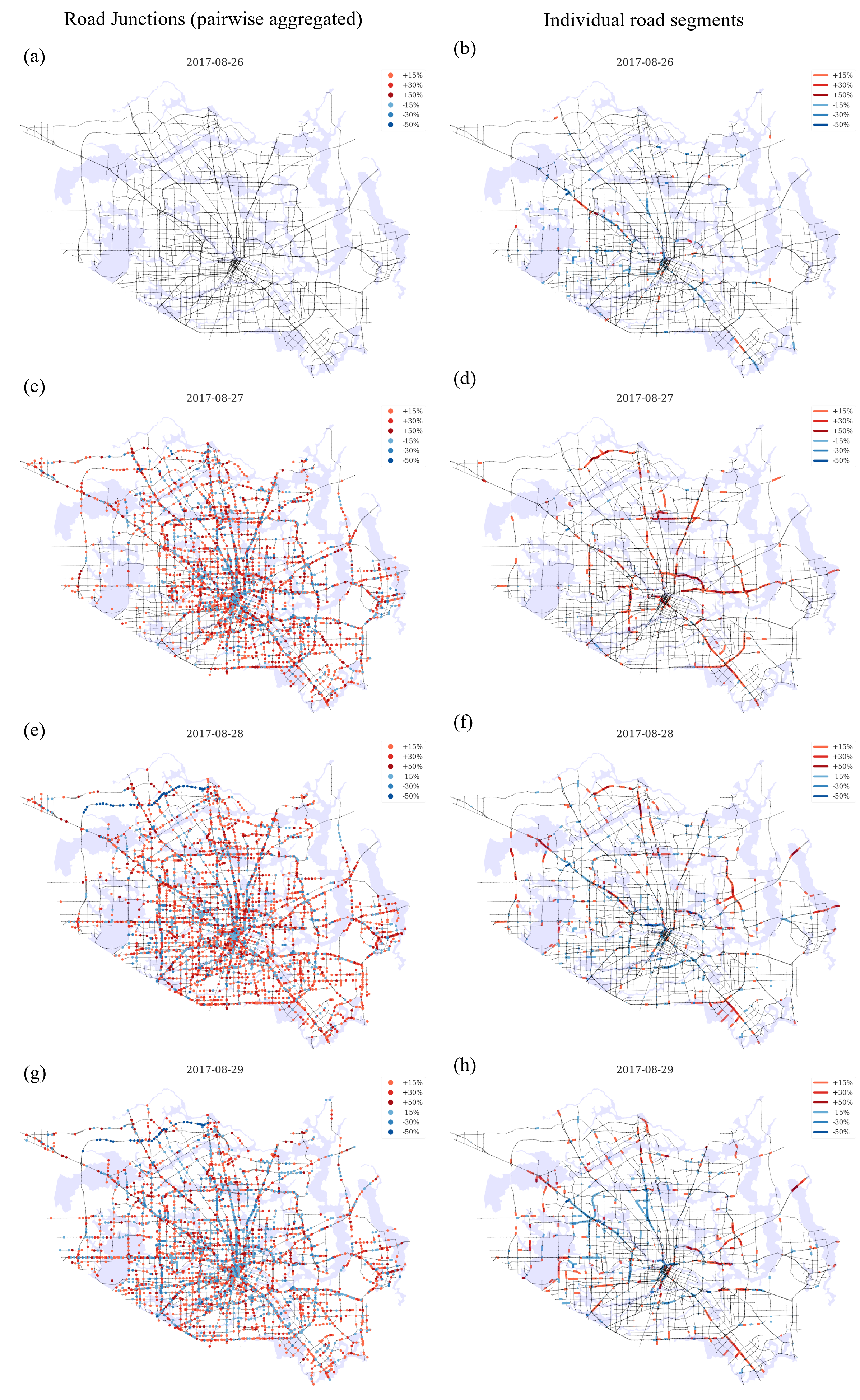}
\caption{Average change of travel time observed in road junctions in comparison to change in travel time for road segments during Harvey. (a), (c), (e) and (d) show the change in travel time for road junctions, and (b), (d), (f) and (h) show the change in travel time for edges for the respective days.}
\label{fig:spatial_change_disaster}
\end{figure*}

To evaluate the dynamics of change in the geometry of traffic networks, we employed the framework shown in Fig. \ref{fig:methods}. First, we processed the raw data to correct any rounding errors and obtained the required temporal resolution. Then we form a spatio-temporal network where edge attributes change with time. We then obtained a pair-wise node distance matrix which were used in the two main approaches used in this study to examine effects of disaster in spatio-temporal traffic networks. Each step in the method is explained below.

\subsection{Data and Pre-processing}

We used a weighted road transportation network of Harris County, Texas, a period before and after Hurricane Harvey flooding (August 1 through September 30, 2017) for this analysis. From INRIX, a private analytics company, we acquired two components of traffic data: road segment location data and segment attributes. The dataset includes the travel time value for each major road segment in Harris County, within which is located the city of Houston. INRIX collects location-based data from both sensors and vehicles. INRIX traffic data contains the average traffic speed of each road segment at 5-minute intervals and their corresponding historical average traffic speed. Each road segment’s geometric information, such as name, geographic locations defining its start and end coordinates, and length, is also available from the INRIX data set. 

Location attributes of some of the road segments varied at the fifth decimal level when taken in degree decimal format. This resulted in some of the road segments being disconnected from the main network, although they were physically connected. To address this, coordinates were rounded off to the fourth decimal to ensure that road segments connect entirely when a network is formed. We then aggregated the attribute information of the road segments at daily resolution to reduce computational effort and provided an overall travel characteristic for the entire day that may differ during rush hours and early morning. Travel time for a road segment was calculated by taking the mean value for all 15-minute intervals for an entire day. 

\subsection{Network construction} 

We constructed a network from the processed road segment data that contains 17,089 edges and 13,550 nodes. Where edges correspond to road segments and nodes corresponds to road junctions. We map each of the road segments based on their location attributes to form this network. The original network consisted of 19712 edges and 15390 nodes but we filtered the nodes and edges from largest connected component in the network and removed some of the nodes that had no data even during non impact days. This step ensured that shortest paths exist between every pairwise junctions in the network as it is primary step in data processing in this paper. Having disconnected nodes or clusters would lead to non-reachable junctions that is not desirable for this analysis. The resulting giant component (largest connected component) accounted for 88\% of the nodes and 87\% of the edges from the original network.

We use this network as a skeleton and construct weighted temporal traffic networks for each of the days from August 1, 2017 through September 30, 2017. 
We use travel time in minutes as edge weight in the network that represents the time for a vehicle passing through an edge (road segment) to traverse though it. 

\subsection{Distance Matrix}

After obtaining temporal networks with travel time as edge attributes, we computed a matrix $A_{13500\times13500}$ where $A_{ij}$ corresponds the shortest travel time from road junction $i$ to road junction $j$ in minutes. Since we treat the transportation network as an undirected graph, travel time from $i \leftarrow j$ is the same as from $j \leftarrow i$, thus yielding a symmetric matrix, where $A_{ij} = A_{ji}$. This distance matrix, where distance between junctions $i$ and $j$ (or $j$ and $i$), is evaluated in time domain. This matrix contains information about the travel time between any junction pairs in the network and collectively represents the travel characteristics of the Harris County traffic network. We use Bellman-Ford algorithm \cite{bellman_1958} to compute the travel time for the shortest paths between every pairwise junction. Since, our network has roughly 13,500 nodes and 17,000 edges, it is computationally expensive to compute the shortest paths between every pairwise nodes in the network. Python natively uses single core for computation so Python libraries such as \textit{swifter, dask}, and native libraries that allow multi-core processing were adopted to speed the computation.  

\subsection{Shortest Paths Analysis}

We use the distance matrix to evaluate the effective spatial transformation of traffic network in Harris County. As the travel time changes for each road segment, the shortest paths between pairwise junctions (nodes) denote the spatial proximity of these junctions in the time domain. Fig. \ref{fig:multilayer} (a) illustrates a sample traffic network showing pairwise travel time with impact on delays due to disasters. Each road segment undergoes a change in travel time during disruption. This could be both positive or negative. If a road segment experiences disruption due to inundation, debris, or other disaster-related obstruction, it would experience increased travel time. This would have compounding effect on travel times between different junctions, as multiple road segments in the path experience disruptions. Other road segments that are not in proximity to damaged areas may experience higher than usual travel times, as they absorb additional traffic routed though them.    

To assess the impact of urban flooding on the entire traffic network, we compute two parameters:  the impact of flooding on the average travel time between every pairwise road junction and the impact of Harvey on different travel time ranges of 15-minute intervals. For both these parameters, the first two weeks of August 2017 was used as baseline to compute change during Harvey. The same days of the week are compared to one another to account for different mobility patterns during different days, such as weekday-weekend effect ~\cite{8316783, weekday_weekend_Katarzyna}. The first parameter provides an idea on the extent of the impact of flooded roads  on the average state of the entire network. The second parameter informs us about the disproportionate impacts on different travel time ranges. 
 
\subsection{Higher Network Dimension Analysis}

The simple network based measures, such as average path length, giant component size in disrupted network, and other network-related measures are not able to fully capture the underlying changes in the network geometry ~\cite{motif_asim_yulia}. Study of interactions between higher-order network features gives a more thorough understanding of topological changes in the network that may uncover important roles that higher-order networks might play in the understanding of dynamics of network topology during disruptions. We capture these hidden dynamics by considering the most basic higherorder feature computed using Betti number of zeroth order (Betti-0) that gives a count of number of connected components at different distance thresholds ~\cite{Torres_2020, asim_yulia_grid}. The Betti numbers are fundamental topological invariants that characterize higher-order networks represented by simplicial complexes ~\cite{bianconi_2021}. 

Let $G=(V, E,\omega)$, an (edge)-weighted graph, be a representation of a temporal traffic network. If we select a certain threshold (or scale) $\epsilon_j>0$ and keep only edges with weights between nodes $u$ and $v$, $\omega_{uv}$, is less than $\epsilon_j$, we obtain a graph $G_j$ with an associated adjacency matrix $A_{uv}=\mathbbm{1}_{\omega_{uv}\leq \epsilon_j}$. Now, changing the threshold values, $\epsilon_1<\epsilon_2<\ldots<\epsilon_n$, results in a hierarchical nested sequence of graphs $G_1 \subseteq G_2 \subseteq \ldots \subseteq G_n$ that is called as a \textit{network filtration}. These filtration levels are depicted in Fig. \ref{fig:multilayer} (b) for a sample network. Each sequence of graphs represents list of junctions that fall within a specific threshold, where threshold represents travel time. Each threshold of travel time indicates road junctions that are accessible within a temporal distance of threshold. At the lowest threshold (0 minutes), no other junction is accessible, so we have same number of components as nodes or junctions in the network. As the threshold increases, more junctions become reachable to these individual junctions; these are connected to form clusters. As the travel time threshold is increased again, these clusters slowly start merging with other clusters to form a single connected component. When the accessibility to a junction is not broken, the last threshold yields just one large connected component, as all junctions are reachable by one another within this time period.    

During non-impact days, the composition of the number of clusters that get formed at different travel time thresholds changes and may show certain characteristics in network geometry indicating higher order dynamics of traffic networks. These changes may not be apparent with basic network measures. Using Vietoris-Rips (VR) complex ~\cite{Carlsson:2009, Zomorodian:2010,Otter_et_al2017}, one of the most popular Topological Data Analysis filtration methods, we track evolution of topological features such as connected components using Betti numbers at different filtration levels. In our case, the distance measure corresponding to travel time in minutes were for a graph, $G=(V, E,\omega)$; the vertices correspond to road junctions, edges correspond to a link between every junction, and weights account for the travel time between the vertices. 

\section*{Acknowledgments}
This material is based in part upon work supported by the National Science Foundation under CRISP 2.0 Type 2 No. 1832662 grant and the Texas A\&M University X-Grant 699. The authors also would like to acknowledge the data support from INRIX. Any opinions, findings, conclusions, or recommendations expressed in this material are those of the authors and do not necessarily reflect the views of the National Science Foundation, Texas A\&M University, or INRIX.

\section*{Data Availability}
The data used in this study are not publicly available under the legal restrictions of the
data provider. Interested readers can request it from INRIX provided here (https://
inrix.com/products/speed/). This paper used high-resolution traffic data (travel speed on
major roads at 15-minute intervals) of Harris County, Texas, during August to September 2017

\section*{Code Availability}
The code that supports the findings of this study is available from the corresponding author upon request.

\bibliographystyle{unsrt}  
\bibliography{references}  

\section{Appendix}

\begin{figure*}[!ht]
\centering
\includegraphics[width=0.8\textwidth]{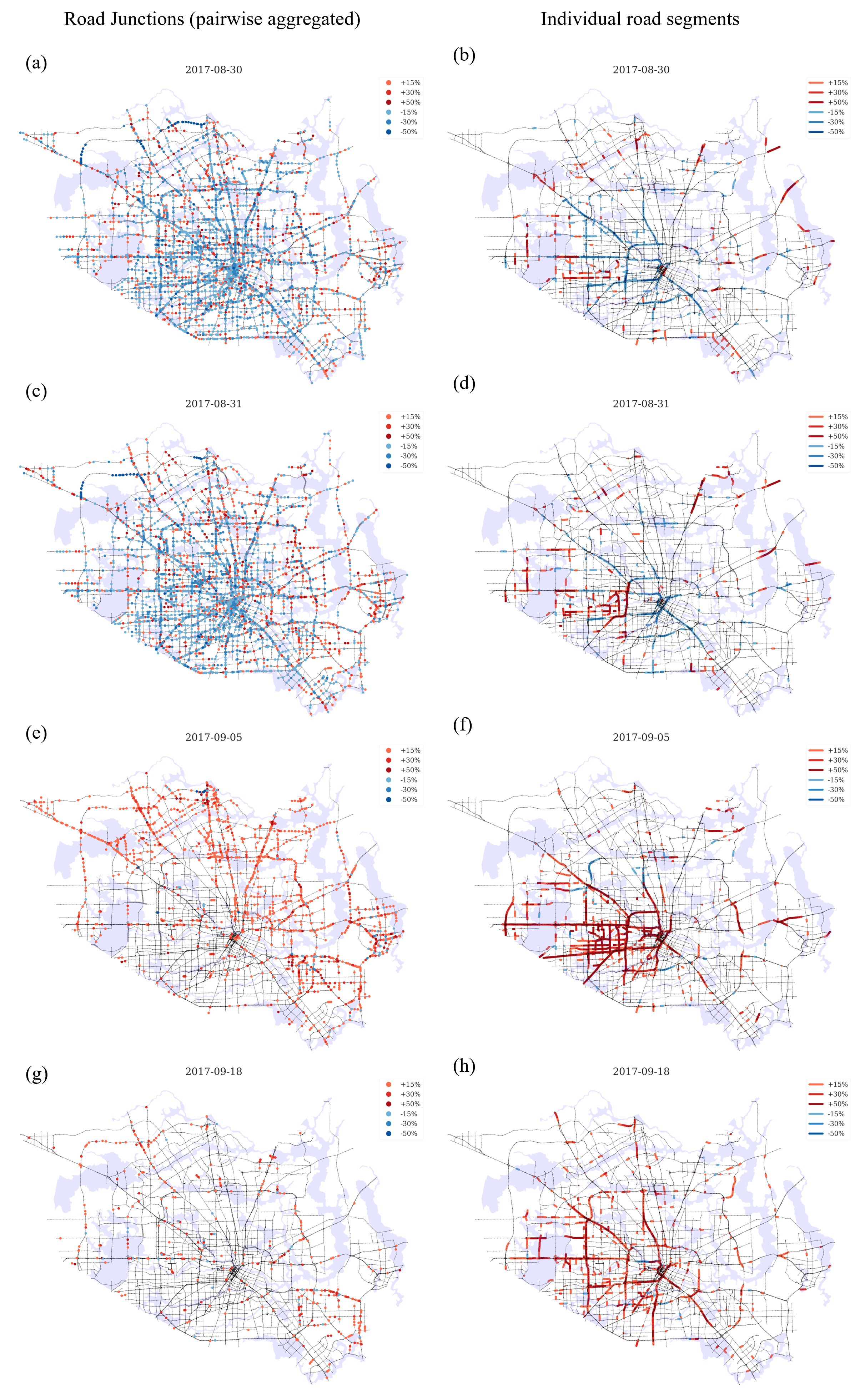}
\caption{Average change of travel time observed in road junctions in comparison to change in travel time for road segments after Harvey. (a), (c), (e) and (g) show the change in travel time for road junctions, and (b), (d), (f) and (h) show the change in travel time for edges for the respective days.}
\label{fig:spatial_change_post_disaster}
\end{figure*}

\end{document}